# The different paths to entropy

L.Benguigui

Solid State Institute   and Physics department
Technion-Israel Institute   of  Technology
32000   Haifa   ISRAEL

In order to understand how the complex concept of entropy emerged, we propose a trip towards the past, reviewing the works of Clausius, Boltzmann, Gibbs and Planck. In particular, since the Gibbs's work is not very well known we present a detailed analysis, recalling the three definitions of the entropy that Gibbs gives. The introduction of the entropy in Quantum Mechanics gives in a compact form all the classical definitions of the entropy. Maybe  one of the most important aspect of the entropy is to see it as a thermodynamic potential like the other thermodynamic potentials, as proposed by Callen. The calculation of the fluctuations of the thermodynamic quantities is thus naturally related to the entropy. We close with some remarks on entropy and irreversibility.

# 1.Introduction

The concept of entropy is not easy to grasp (even for physicists) and frequently entropy is seen as very mysterious quantity. It received a very large number of interpretations, explications, applications. The literature on entropy is huge and it is possible in any work on entropy to mention only a very small fraction of the already published works on the subject [1]. Only for the curiosity of the reader one can mention that not all these applications have a real rational basis connected firmly with physical laws [2].

The difficulty to understand the concept of entropy has a long history as one has testimonies from the beginning of the twentieth century until today One can quote James Swinburne who initiated with John Perry a debate on entropy from 1902 to 1907 [3]. It results a series of articles in electrical engineering journals. Swinburne himself wrote a book on Thermodynamics: *Entropy or Thermodynamics from an Engineer's Standpoint and the Reversibility of Thermodynamics.* Swinburne wrote:

*As a young man I tried to read thermodynamics, but I always came up against entropy as a brick wall that stopped my further progress. I found the ordinary mathematical explanation, of course, but no sort of physical idea underlying it. No author seemed even to try to give any physical idea. Having in those days great respect for textbooks, I concluded that the physical meaning must be so obvious that it needs no explanation, and that I was especially stupid on the particular subject.*

Recently Ben Naim [4] published three books on entropy with the explicit goal to make understanding entropy easy for all. The simple fact that he needs three books for this is a clear sign of the complexity and the difficulty of this task.

In this paper, I propose a kind of trip through time, coming back to the original works on entropy. It may be illuminating to analyze how the concept was born and how it evolved through the works of Clausius, Boltzmann, Gibbs and Planck. There are already several works on the history of entropy (see for example ref.5) but it appears that they are far from being complete: very often only the works of Clausius and Boltzmann are described, and further developments are forgotten.

Firstly I intend to go to the initial appearance of the concept by Clausius in 1864 and afterward to follow two other important developments in the definition of entropy: Boltzmann and Gibbs. Though the fundamental work of Gibbs is frequently mentioned, it remains not very well known. Who does remember that Gibbs proposed three definitions for the entropy? This is the reason why it seemed that there is interest in exposing the Gibbs's work in some detail. But I shall be more concise about Boltzmann who is much well known. On the way I shall also mention Planck and his work on the black body radiation since the use of the Boltzmann entropy was essential for his pioneering work. I shall go on to the definition of entropy in Quantum Mechanics. Finally I shall discuss how the entropy developed in the modern view of Thermodynamics.

## 2.Clausius, the father of the Entropy

Clausius published a series of papers (in German) on the theory of heat and its applications to engines, from 1850 to 1864. The papers were translated to English and published all together in a book in 1872, *The Mechanical Theory of Heat with its Applications to Steam-Engine and to the Physical Properties of Bodies.*[6] The book is divided into 9 memoirs and I shall mention only those which are relevant to the subject of entropy.

To understand the development of the Claudius's ideas, one has to come back to the work of Sadi Carnot *Réflexions sur la puissance du feu et sur les machines propres à développer cette puissance* published in 1824 [7]. The important result of his work is that it is not possible to transform completely into mechanical work a given quantity of heat extracted from a hot reservoir in a cyclic process. A part of the heat must be transferred to a cold heat reservoir. This is true even in a case of an engine working reversibly. In this case, the engine can be used in the reverse direction and a quantity of heat may be transferred from the cold reservoir to the hot reservoir but a mechanical work must be provided to the engine, work which is transformed into heat. Clausius formulated the Second law of Thermodynamics in saying that heat cannot pass from a cold reservoir to a hot one by itself.

Thus the Second law of the Thermodynamic is simply the impossibility to transform heat completely into mechanical work in a cyclic process. The impossibility for heat to pass spontaneously from a cold reservoir to a hot reservoir can be deduced from this formulation of the Second Law. We have to add that it is not possible to make an analogy with a water fall since the heat is not a conserved quantity but a form of energy.

We note two things. Apparently, this formulation based on the Carnot engine or the Carnot cycle does not introduce the notion of non-reversible process since the engine can work in the two directions. Secondly, this formulation does not introduce entropy.

However, irreversibility is hidden in that there is asymmetry in this formulation of the Second Law: in a cyclic process, it is possible to transform completely work into heat when the contrary is not possible. Moreover, heat passes spontaneously from a hot reservoir to a cold reservoir while the contrary is possible only in investing work, i.e. non-spontaneously. The spontaneous process is therefore irreversible.

In the fourth memoir (published in 1854), Clausius proceeds to an original analysis of the Carnot cycle and sees in it two *transformations*: the first is the transfer of heat from the cold to the hot reservoir (or from the hot to the cold) and the second is the transformation of work into heat (or heat into work). He considers these two *transformations* as equivalent, not in the meaning that one can replace the other but rather that one *transformation* cannot appear without the other. Now he proposes to introduce a new concept that he calls the *equivalent-value* of these *transformations*. If Q is the heat transformed in work (or the work into the heat Q) from the hot reservoir at temperature $T_h$, the *equivalent-value* is

$$(Q / T_h) \qquad (1)$$

Now if Q' is the heat transferred from one reservoir to the other (the cold one having a temperature $T_c$) the *equivalent-value* is

$$Q' (1/T_h - 1/T_c) \qquad (2)$$

Following Clausius, since these *transformations* are equivalent their *equivalent-values* are equal. Supposing that the Carnot cycle is working as an engine giving work to the surrounding, one has Q > 0 and Q' < 0. Equating (1) and (2) give the well-known relationship between the heat expelled from the hot reservoir (Q – Q'), that received by the cold reservoir (– Q') and the temperatures of the reservoirs

$$(Q - Q') / T_h + Q' / T_c = 0 \qquad (3)$$

Generalizing to a cyclic process as the limit of infinity of elementary Carnot cycles, Clausius gets that in a reversible cyclic process, one has

$$\int dQ/T = 0 \qquad (4)$$

the integral being calculated along the cycle. Or, in other words, the differential

$$dS = dQ / T \qquad (5)$$

is an exact differential and by this way Clausius introduces a new variable S which depends only on the state of the system. In fact by this way one can only calculate the difference of S between two different states (A and B) of the system as

$$S_B - S_A = \int_A^B dQ/T \qquad (6)$$

following a reversible path

But it is remarkable that at this stage he does give the name "entropy" to this new variable.

Considering an irreversible cyclic process, Clausius finds that the *equivalent values* of *the two transformations* are not equal but the positive *transformation* dominates. About a irreversible cyclic process, he finds that

$$\int dQ/T \geq 0 \qquad (7)$$

Here also the integral is calculated along the cycle.

In his sixth memoir (published in 1862) Clausius looks for a physical meaning of his new variable S and he introduces the concept of *disgregation* or the magnitude of the degree in which molecules of a body are separated from each other. This memoir presents a long discussion about the relation between disgregation and the variable S. This aspect of Claudius's work has only a historical interest but this shows that, at this time, the connection of the new variable S with irreversibility was not clearly established for him.

The ninth and last memoir (published in 1865) is almost completely devoted to thermodynamics calculations using the two Laws of Thermodynamics (or the two *theorems* as Clausius calls them) as one can find them in many textbooks. Almost at the end of his book, at the page 357 (of a book of 374 pages) he coins the term "entropy" and finds some new thermodynamics relationship. On page 362, he considers the case of irreversible processes and he gets (but in a different form) the famous expression relating entropy variation along an irreversible process (not cyclic) and the integral $\int dQ/T$

$$S - S_0 \geq \int_{irr} dQ/T \qquad (8)$$

In an adiabatic irreversible process dQ = 0 and one has the fundamental relation between entropy and irreversibility:

**In a closed system (adiabatic transformation) the entropy cannot decrease, it remains constant or increases**.

Suddenly (two pages before the end of his text), Clausius (under the influence of british scientists like Thomson and Rankine as he admits) states that

*"if all the changes of condition occurring in universe the transformations in one definite direction exceed in magnitude those in the opposite direction, the entire condition of universe must always continue to change in that first direction, and the universe must consequently approach incessantly a limiting condition,"*

He is clearly less pessimistic than Thomson [8] who predicts in 1852 that

*"Within a finite period of time past, the earth must have been, and within a finite period of time to come the earth must again be, unfit for the habitation of man as at present constituted.."*

It is likely that Clausius "discovers" the connection between entropy and irreversibility only relatively late during the development of his works on heat. Curiously he never asked a question about the origin of irreversible process. But he found a new function: the entropy as a state function depending only on the parameters defining the system

The problem of irreversibility or how to explain the origin of irreversible processes becomes a subject of an intense debate among scientists in the second part of the $19^{th}$ century and Ludwig Boltzmann made important contribution to this subject.

## 3. Boltzmann and the famous formula $S = k_B \ln W$

The first contribution of Boltzmann to the problem of irreversibility is that he called the H-Theorem. With this theorem (published in 1872), Boltzmann claims that he have got a general theorem which can be seen as a proof of the Second law of the Thermodynamics and gives an explanation of irreversibility. It is a highly controversial paper and I do not bring here the different points of view about this theorem [9].

To prove this theorem, Boltzmann first introduced in a paper (also published in 1872) his equation called today by his name: "the Boltzmann equation". It is an integro-differential equation concerning the distribution function f(**v**,t) such that f(**v**,t)d$^3$**v** gives the relative number of molecules of a gas between **v** and **v** + d$^3$**v** at time t (**v** being the vector velocity). The H-theorem states that the quantity defined as

$$H = \int f(\mathbf{v},t) \ln f(\mathbf{v},t) \, d^3\mathbf{v} \qquad (9)$$

decreases with time or

$$dH/dt \leq 0 \qquad (10)$$

until that dH/dt = 0 at equilibrium.

It is tempting to associate the quantity H with −S as Boltzmann wrote (quoted in reference 10 )

*Thus, one may prove that, because of the atomic movement in systems consisting of arbitrarily many material points, there always exists a quantity which, due to these atomic movements, cannot increase, and this quantity agrees, up to a constant factor, exactly with the value that I found in [a paper published in 1871] for the well-known integral $\int dQ/T$.*

The rationale of Boltzmann was to find a quantity which reaches an extremum in the equilibrium state. However, as mentioned above, the H-theorem was severely criticized and until today it is object of discussions [9]. Contrarily to the Boltzmann's opinion, −H is not considered as the entropy.

The second important contribution of Boltzmann concerns his famous formula S = $k_B$ Ln W based on his H theorem, in a paper published in 1877. I shall survey the essential argument of this article and shall not enter into the details of the derivation. The principal point is the definition of a microstate. A microstate corresponds to a macroscopic state of the system (in equilibrium or not) and it is defined by the positions and the momenta of all the particles of the system. Clearly a large number of microstates corresponds to this particular macroscopic state. One can imagine that following the time the system passes from one microstate to another. A fundamental hypothesis concerning a closed system is that the probability of occurrence of a particular microstate is identical for all the microstates of the system.

Consequently the equilibrium state is that with the largest possible number of microstates since it corresponds to the largest probability for the system itself.

If a closed system is not in an equilibrium state, it will evolve to states with larger and larger number of microstates until it reaches the equilibrium state with the largest number of microstates corresponding to its energy. But one has to refine this picture: even being in the equilibrium state, the system may pass to another macroscopic state with a smaller number of microstates. However, the probability that the system reaches such a state is very, very low such it is never observed. Here is the important point: the introduction of the notion of the probability for the occurrence of a state of the system. As mentioned above the equilibrium state is the one with the largest probability and once it reaches this state it goes backward only with a very, very small probability. (A more detailed analysis shows that some fluctuations around the equilibrium state exist but one does not consider them for the time. See below)

The following step is to identify the largest probability with entropy and to write the proportionality of S with the logarithm of the number W of microstates

$$S \approx \text{Ln } W \qquad (11)$$

The Second Law of the Thermodynamics takes now a probabilistic character. The irreversibility appearing when a system out of equilibrium comes back to equilibrium is explained in terms of probability which increases. Clearly, it is not the end of the story but an important step is made. For details about the works of Boltzmann see refs. [10] and [11].

But until now an important aspect of the problem was neglected. We have admitted as evident that microstates can be distinguished. How is it possible to differentiate microstates? What is the parameter which defines different microstates? There is no problem if the different microstates are identified by means of discrete variables. It is not possible to consider different microstates if this parameter is continuous (like energy for example) since, even in a limited range of the energy, there is an infinity of different microstates, each with a different value of the energy. The solution is to make the energy a discrete quantity, as effectively Boltzmann did, charaterized by energy steps $\Delta$. Two microstates are called different if their energies are different at least by $\Delta$. It results that the number of microstates

W and consequently the entropy depend on the choice of Δ. In fact since the entropy is the logarithm of W, it results that the entropy is defined with an arbitrary constant.

However something is not self-evident. In the case of particles with continuous values of the coordinates indicating their continuous values of their positions and of their momenta, is the process of discretization permitted? In other words, in a world of continuous variables (coordinate, momentum, energy etc.) is it possible to calculate quantity like entropy by discretization?

The question is not futile and we shall come back to this question below with Planck and his analysis of the black body radiation. In fact the famous formula (11) can be used only in a world of discrete quantities. It is effectively the world of Quantum Mechanics where, in a limited volume, the relevant quantities are discrete.

4.<u>Planck, the entropy and the law of the black body emission</u>

It is well known that Planck succeeded in finding the correct expression for the density of energy for the black body emission by discretization of the energy. He uses the method known as the microcanonical ensemble [12]. In this method one calculates the entropy of the system with a given value of the energy using the Boltzmann formula $S \approx Ln\ W$. From the relation $(\partial S/\partial E) = 1/T$ one can deduce the dependence of the energy on T and consequently the dependence of the entropy on T. It is exactly what Planck did in supposing that the energy of one resonator of the black body can take only discrete values and he was able to calculate the number of possible states with a given energy. As known this paved the way to the development of Quantum Mechanics.

It is clear that Planck needed this discretization in order to get the good result. One can ask an apparently innocent question   In order to emphasize the influence of the discretization, consider the following problem: N independent particles can take energies form 0 to ∞, in contact with a thermal reservoir at temperature T. What are the energy and the entropy of the system as functions of T? (Here we do not introduce a second variable as usual like the volume . We suppose that the volume is constant). We shall do that following two ways. In the first the energy is continuous as proposed. In the second, we shall suppose that the particle energy can be divided into discrete quantities $E = Kn$ with $n = 0, 1, 2, 3 ,,,,$ and K is a constant. We shall calculate E and S as functions of T and K and after wad we shall do K→0 to recover the continuity of the individual energies.

In the first case the total energy of the system is

$$E = N \int_0^\infty E \, \exp\left(-\frac{E}{kT}\right) dE \Big/ \int_0^\infty \exp\left(-\frac{E}{kT}\right) dE \tag{12}$$

or

$$E = N \, k_B \, T \tag{13}$$

The entropy can be calculated from the expression $(\partial S/\partial T) = (1/T)(\partial E/\partial T)$ or

$$S = \int \left(\frac{1}{T}\right)\left(\frac{\partial E}{\partial T}\right) dT \tag{14}$$

One gets

$$S = N \, k_B \, \mathrm{Ln}(T/T_0) \tag{15}$$

It is necessary to introduce an integration constant, the temperature $T_0$, since T cannot go to zero (this gives $S \to \infty$). The expression (15) has meaning only for $T > T_0$. However it is always possible to calculate the entropy difference $\Delta S$ between two states at temperatures $T_1$ and $T_2$. One has $\Delta S = N \, k_B \, \mathrm{Ln}(T_2/T_1)$.

In the second method, we calculate the one particle partition function Z:

$$Z = \sum \exp(-E/kT) \tag{16}$$

or

$$Z = \sum_0^\infty \exp\left(-\frac{nK}{kT}\right) = \frac{1}{1-\exp(-\frac{K}{kT})} \tag{17}$$

The average energy of the system is

$$E = N \, [\partial(\mathrm{Ln}Z)/\partial \beta] \quad (\beta = 1/k_B T) \tag{18}$$

$$E = \frac{KN}{\exp\left(\frac{K}{kT}\right) - 1} \tag{19}$$

The entropy is obtained from $S = -(\partial F/\partial T)$ where F is the free energy

$$F = -N\, k_B\, T\, \text{Ln}\, Z \tag{20}$$

$$F = N\, k_B\, T\, \text{Ln}[1 - \exp(-K/k_B T)] \tag{21}$$

And finally

$$S = -N\, k_B\, \text{Ln}[1 - \exp(-K/k_B T)] - N(k_B K/T)\, \frac{1}{\exp\left(\frac{K}{kT}\right) - 1} \tag{22}$$

It is important to note that this expression is correct at all temperatures and in particular S tends to zero if T decreases towards zero.

Now we shall let K going to zero. For the energy we get $E = N\, k_B\, T$ as above (13). But for the entropy the things are different. For $K \to 0$, one has

$$S(K \to 0) = k_B\, N\, [\text{Ln}\, (k_B\, T\, /\, K) - 1)] \tag{23}$$

In other word there is no limit for S when $K \to 0$. However the entropy difference is as above $\Delta S = N\, k_B\, \text{Ln}(T_2/T_1)$. One sees that the limit $K \to 0$ corresponds to the classical limit at high temperature.

This simple calculation gives us an important lesson. With the help of the entropy (and not the energy) we discover that at low temperature, the energy of particles must be discrete: if not the entropy cannot be calculated at low temperature (see 15 and 23).

The conclusion is that the entropy is the gate to the world of discreteness or Quantum Mechanics.

## 5. Gibbs and Statistical Mechanics

In a small book entitled *Elementary Principles in Statistical Mechanics* published in 1904 Gibbs presents the basis for what is called "Statistical Mechanics" or how to calculate macroscopic properties of a huge number of particles [13]. Since it impossible to follow the motion of each particle because of their huge number, what is it possible to tell about this group of particles knowing that macroscopic properties include thermal phenomena ? In other words, Gibbs intends to give a rigorous basis for Thermodynamics as indicated by the second title of his book: *The Rational Foundation of Thermodynamics.*

Wishing to find the thermal properties of a particular system of a large number of particles, Gibbs uses the trick of an *ensemble.* He considers a very large number of identical systems, the ensemble and supposes that each system in the ensemble is in a particular situation i.e. that in each system the particles are in different positions with different momenta or in a different microstate. Gibbs defines the density D in the extension in phase (following the words of Gibbs) or in the phase space with dimension 2n = 6 times the number of particles in one system as follows. It is the number of systems having the position of their particles described by the coordinates $q_1, q_2, q_3 \ldots q_i \ldots q_n$, and momenta $p_1, p_2, p_3 \ldots p_i \ldots p_n$ in a small box (size given by $dq_1 dq_2 \ldots dq_i \ldots dq_n \, dp_1 dp_2 dp_1 dp_2 \ldots dp_i \ldots dp_n$) around these positions and momenta values. One already sees the need to make a discretization of the quantity D since the quantities dq's and dp's are very small but have necessarily some finite values.

Gibbs gives a demonstration of the well-known *Liouville* theorem which states that if the forces acting on the particles of one system are function only of the coordinates (with or without the time), the density-in-phase D is constant in time. Instead of using D, Gibbs defines the probability to find one system in a given box of the phase space by P = D / N where N is the number of systems in the ensemble. From the knowledge of P, it is possible to calculate all the average or macroscopic properties of the system. We shall call P the distribution probability. One notes that this definition of P does not imply that the systems are in equilibrium and thus one can define P also for situations out of equilibrium.

Gibbs considers what he calls the "canonical ensemble" i.e ensemble of identical systems controlled by the temperature and other external parameters such as the volume. The differential of the work made by these external parameters is $\Sigma - A_i \, da_i$. For this ensemble Gibbs proposes (apparently without justification, but one can see that as a hypothesis) that

$$P = \exp[(\Psi - \varepsilon)/\Theta] \qquad (24)$$

where $\Psi$ and $\Theta$ are constants and $\varepsilon$ is the energy of one system. Since one system is controlled by the temperature and external parameters, the energy is not constant but fluctuates. P dε gives the distribution of one system i.e. the probability to find a system with energy between $\varepsilon$ and $\varepsilon + d\varepsilon$. The logarithm of P is linear with the energy

$$\eta = \operatorname{Ln} P = (\Psi - \varepsilon)/\Theta \qquad (25)$$

Gibbs calls the quantity $\eta$ index of probability. By definition of P, one has

$$\int P \, dq_1\ldots dq_n dp_1\ldots dp_n = 1 \qquad (26)$$

$$\int \exp[(\Psi - \varepsilon)/\Theta] \, dq_1\ldots dq_n dp_1\ldots dp_n = 1 \qquad (27)$$

and the average value of a quantity B is given by

$$B_{av} = \int B \, P \, dq_1\ldots dp_n = \int B \exp[(\Psi - \varepsilon)/\Theta] \, dq_1\ldots dq_n dp_1\ldots dp_n \qquad (28)$$

### 5.1 The first definition of entropy

Differentiating the following expression (relatively to $\Theta$), obtained from (27)

$$\exp(-\Psi/\Theta) = \int \exp(-\varepsilon/\Theta) \, dq_1\ldots dq_n dp_1\ldots dp_n \qquad (29)$$

and taking the averages, Gibbs gets (see details in Appendix A)

$$d\varepsilon_{av} = -\Theta \, d\eta_{av} - \Sigma \, (A_i)_{av} \, da_i \qquad (30)$$

with

$$\eta_{av} = \int \eta \, P \, dq_1\ldots dp_n = \int P \operatorname{Ln} P \, dq_1\ldots dp_n. \qquad (31)$$

Equation (30) is similar to the well-known thermodynamic equation giving the differential of the internal energy U

$$dU = T \, dS - \Sigma \, A_i \, da_i$$

Thus Gibbs concludes that the parameter $\Theta$ is proportional to the temperature and that the entropy is proportional to

$$S \approx -\eta_{av} = -\int P \ln P \, dq_1 \ldots dp_n . \qquad (32)$$

In other words the entropy is the average of the logarithm of the distribution described by P (with a change in the sign in order to make the entropy positive) [p.45, ref.13 ]. This definition is general even if P is different from the expression (24).

It is the most important expression of the entropy and it seems that Gibbs was the first to find it.

5.2 The second definition of entropy

For the second definition of the entropy, Gibbs introduced the function $V(\mathcal{E})$ defined as the "volume" of the phase space with energy smaller than a given value of the energy. In other words, in the framework of the canonical ensemble, V is the number of boxes in the phase space (i.e in which there are coordinates and momenta of at least one system) for which the total energy is less or equal than a given value of energy:

$$V = \int dq_1 \ldots dq_n dp_1 \ldots dp_n \qquad (33)$$

the integral being calculated for values of $q_1$, $q_2$ ….$q_n$, $p_1$, $p_2$ …..$p_n$ corresponding to energies smaller than a given value $\mathcal{E}$.

The derivative $(dV/d\mathcal{E})$ can be seen as the energy density of the space phase since the quantity $(dV/d\mathcal{E})d\mathcal{E}$ indicates how many boxes there are with energy between $\mathcal{E}$ and $\mathcal{E} + d\mathcal{E}$. It results now that in the above integrals the differential form $dq_1\ldots\ldots dp_n$ can be replace by $(dV/d\mathcal{E})d\mathcal{E}$. Thus the integral (28) giving the average value of the quantity B is now

$$B_{av} = \int B \exp[(\Psi - \mathcal{E})/\Theta] \, (dV/d\mathcal{E})d\mathcal{E} \qquad (34)$$

A new function is now defined

$$\Phi = \ln (dV/d\mathcal{E}) \qquad (35)$$

and Gibbs shows two following important results (see Appendix B for details of the calculations);

1. The average of the derivative $(d\Phi/d\varepsilon)_{av}$ is equal to $1/\Theta$.

2. The value of $(d\Phi/d\varepsilon)$ for the most probable value of the energy $\varepsilon_0$ is also equal to $1/\Theta$.

Consequently, since $\Theta$ was already identified with the temperature, Gibbs concludes that $\Phi_o$, the value of $\Phi$ for the most probable value of the energy, is proportional to the entropy:

$$\Phi_0 = [Ln(dV/d\varepsilon)]_0 = \text{entropy} \qquad (36)$$

In the case of the of the microcanonical ensemble (with a definite of the energy) the quantity $(dV/d\varepsilon)d\varepsilon$ calculated for this specific value of the energy is the "volume" of the shell of thickness $d\varepsilon$. More simply $(dV/d\varepsilon)$ is the "surface" of the "volume" $V(\varepsilon)$ and it gives the number of microstates.

One can see that the second definition (for the microcanonical ensemble) of the entropy is equivalent to that of Boltzmann $S = Ln\ W$. W is the largest possible number of microstates with a definite energy (or the energy with the largest probability). The corresponding number in the Gibbs's theory is in fact, as seen above, $(dV/d\varepsilon)d\varepsilon$ and its log is $Ln[(dV/d\varepsilon) + Ln(d\varepsilon) = LnW$. Thus the two expressions are different by a constant and this shows their equivalence. In other words

$$S(Boltzmann) = S(Gibbs) + \text{constant}.$$

However, the definition (36) corresponds to the macrocanonical ensemble and one concludes that the average of the "surface" of the "volume" $V(\varepsilon)$ is the entropy and is equal to its value for a given energy in the microcanonical ensemble.

The procedure of Gibbs has the advantage to be valid even with continuous quantities i.e. at not too low temperatures.

5.3 The third definition of entropy

Gibbs gets the following result (see Appendix C):

$$[d\mathcal{E}/d(\ln V)]_{av} = \Theta \qquad (37).$$

Consequently $(\text{Ln}V)_{av}$ is also a candidate for the entropy. Since $[\exp(-\Phi)V]_{av} = \Theta$, Gibbs can write the differential of the averaged energy as

$$d\mathcal{E}_{av} = [\exp(-\Phi)V]_{av}\, d(\text{Ln } V)_{av} - \Sigma\, (A_i)_{av}\, da_i \qquad (38)$$

Why the mean values of Ln V and of Ln (dV/d$\mathcal{E}$) are both proportional to the entropy (with may be different constants)? To understand this point one considers an ideal gas of n atoms with mass m. To calculate $V(\mathcal{E})$, one begins to calculate $V(p)$ where p is the absolute value of the linear momentum. The "volume" limited by $V(p)$ is the "volume" of the hypersphere of radius p in 3n dimensions

$$V(p) = p^{3n}\, (\pi)^{3n/2}/\Gamma(3n/2 + 1) \qquad (39)$$

or with $\mathcal{E} = p^2/2m$

$$V(\mathcal{E}) = (\mathcal{E})^{3n/2}(2m\pi)^{3n/2}/\Gamma(3n/2 + 1) \qquad (40)$$

Now one has Ln V proportional to $(3n/2)$ Ln $\mathcal{E}$ and dV/d$\mathcal{E}$ proportional to $(3n/2 - 1)$Ln $\mathcal{E}$. If n is much larger than 1, the dependence of Ln $\mathcal{E}$ and that of dV/d$\mathcal{E}$ with $\mathcal{E}$ is the same. Thus the means of Ln $\mathcal{E}$ and dV/d$\mathcal{E}$ differ only by some constants.

In the chapter XIV of his book, Gibbs discusses the different forms of the entropy: $\eta_{av}$, $\Phi$ and Ln V. The important point is that the three expressions are equal in the limit of an infinite number of degrees of freedom i.e. an infinite number of particles. He discusses what could be the best expression for the entropy and he hesitates between the second or the third definition.

5.4 States out of equilibrium and irreversibility

A large part of Gibbs's book is devoted to equilibrium states but he envisages also situations out of equilibrium, when the distribution probability P is not equal to that of equilibrium. In particular he demonstrates a series of inequalities [chapter XI] which show the particular character of the entropy as a quantity which in certain circumstances can only increases or dos not change.

An important case is given by the theorem I of chapter XI. If an ensemble of systems has its distribution probability P different from that of the canonical ensemble i.e. that given by (12) above, the quantity $\eta_{av}$ is larger than for the canonical ensemble. An ensemble with a distribution probability different from that of the canonical is not in equilibrium. Thus when the ensemble comes back to the equilibrium, its $\eta_{av}$ is less than that of the ensemble out of equilibrium. Since $\eta_{av}$ is proportional to $-S$, this implies that when the ensemble comes back to equilibrium, its entropy increases.

The importance of the work of Gibbs lies in the fact that it is rigorous and there is only one hypothesis: that the quantity $\eta$ is linear in energy. Through different ways, Gibbs gets the fundamental thermodynamics law $dU = T\, dS - \Sigma\, A_i\, da_i$ and the properties of the entropy, and this justifies the hypothesis. Furthermore, Gibbs gives the equivalence of the different expressions of the entropy.

The introduction of the entropy corresponds to the fact that our knowledge of a system is limited and can be only statistics. It is the main lesson of the approach of Gibbs besides its rigorous aspect. The problem of irreversibility is evocated and is discussed in relation with probability. Reversibility is a mathematical property of the equations of mechanics but the real world is probabilistic. Distinction between past and future can be made as it is expressed in this very suggestive remark of Gibbs at the end of the chapter XII:

*But with the distinction of prior and subsequent events may be immaterial with respect to mathematical fictions, it is quite otherwise with regards to the events of the real world. It should not forgotten, when one ensemble is chosen to illustrate the probabilities of events in the real world, that while the probabilities of subsequent events may often be determined from the probabilities of prior events, it is rarely the case that probabilities of prior events can be determined from those of subsequent events, for we are rarely justified to excluding the consideration of the antecedent probability of the prior events.*

In other words an arrow of time is present in the real word due to its probabilistic character.

To conclude this survey of the Gibbs's work in Statistical Mechanics, one has to emphasize that the starting point of Gibbs is not to look for a quantity which must be maximum at the equilibrium state. In other words the problem of the irreversibility is not his starting point but only a consequence of his analysis.

6. Entropy in Quantum Mechanics

The well-known expression for the entropy in Quantum Mechanics is

$$S = -k_B \sum \rho_n \operatorname{Ln} \rho_n = \operatorname{Tr}[\rho \operatorname{Ln} \rho] \qquad (41)$$

where $\rho$ is the density matrix ($k_B$ is the Boltzmann constant). One notes already the similitude with the Gibbs expression (32). However following Landau and Lifshitz [14], it is not possible to separate the two statistical aspects: that due to Quantum Mechanics and that due to Statistical Mechanics. They introduce a statistical matrix in place of the density matrix.

I think that is important to understand the development of the notion of entropy in Quantum Mechanics and how one can get (41). For that I present the derivation given by Landau and Lifshitz in their book *Statistical Physics* [14]. Interesting enough to see that one recovers all the above expressions but in the frame of the Quantum Mechanics.

Following an idea of Gibbs, one considers a very large system which is closed, i.e. isolated from the external surrounding in equilibrium. Now one divides it (by the thought) into a large number of subsystems with the same volume. They are all identical and large enough that one can see them as macroscopic bodies. It is possible to see these subsystems as a physical realization of a *canonical ensemble* as invented by Gibbs. They have all the same temperature and the same volume but the energy fluctuates from one subsystem to another.

In Quantum Mechanics, one does not follow the individual motion of particles since the state of a system is defined by its wave function and its energy. The wave functions of one of the subsystems are denoted as $\psi_n(q)$ where q indicates the coordinates and the index n the state and $E_n$, the energy of the state

Suppose now that at a given time the subsystem is described completely by the wave function $\Psi$ which can be developed using the functions $\psi_n(q)$ as

$$\Psi = \sum c_n \psi_n(q) \qquad (42)$$

The mean value of a quantity f is given by

$$f_{av} = \sum_n \sum_m c_n^* c_n f_{nm} \qquad (43)$$

where $c_n^*$ is the conjugate of $c_n$ and the $f_{nm}$ are the matrix elements of the operator $f^o$

$$f_{nm} = \int \psi_n^* f \psi_m \, dq \qquad (44)$$

To pass now to the subsystems in a canonical ensemble, one adopts the formalism developed above (expression (43)) and writes

$$f_{av} = \sum_n \sum_m w_{mn} f_{nm} \qquad (45)$$

where now the $w_{mn}$ are the matrix elements of the distribution we look for. They can be seen as the matrix elements of a new operator $w^o$ and the mean of f can now be written as

$$f_{av} = \sum_n (w^o f^o)_{nn} = \mathrm{Tr}\,(w^o f^o) \qquad (46)$$

where in (46) only the diagonal elements are taken. One can show also that

$$w_{nn} = w_n > o \quad \text{and} \quad \mathrm{Tr}(w^o) = \sum_n w_n = 1 \qquad (47)$$

As above one considers, for one subsystem, the distribution probability dependent on the energy $E_n$, $w(E_n)$, from which the mean values of the various important quantities can be calculated.

If one speaks about one subsystem, we can remove the subscript n. If the number of quantum states with energy less than a given value E is described by the function $\Gamma(E)$, the number of states between E and E + dE is given by $(d\Gamma(E)/dE)dE$. Thus the probability that the energy of the system is between the values E and E + dE is the product of the probability $w(E)$ by the number of states with energy between E and E and E + dE

$$W(E)\,dE = [d\Gamma(E)/dE]\,w(E)\,dE \qquad (48)$$

with the condition $\int W(E)dE = 1$.

The function W(E) is in general peak shaped, since the macroscopic value of the energy given by its average $E_{av}$ is also equal to the most probable value $E_m$ or $E_m = E_{av}$.

It results that the function W(E) has values different from zero only in the vicinity of the peak value. This gives the possibility to define a width $\Delta E$ of the function as given by the product

$$W(E_m) \Delta E = 1 \tag{49}$$

where $E_m$ is the value of E at the maximum of W(E). Writing from (48)

$$W(E_m) = [d\Gamma(E_m)/dE] \, w(E_m)$$

$$W(E_m) \Delta E = [d\Gamma(E_m)/dE] w(E_m) \Delta E \tag{50}$$

$$= w(E_m) \Delta\Gamma = 1 \tag{51}$$

where $\Delta\Gamma$ is defined as

$$\Delta\Gamma = [d\Gamma(E_m)/dE] \Delta E = [d\Gamma(E_{av})/dE] \Delta E \tag{52}$$

The derivative $[d\Gamma(E_m)/dE]$ gives the number of states around $E_m$ and $\Delta\Gamma$ is the number of states in the interval $\Delta E$ in which the probability function W(E) has values different from zero. Thus $\Delta\Gamma$ is a measure of the spread of the energies of the microstates of one subsystem around the most probable value. We recall that the most probable value is also the average: $E_m = E_{av}$

The entropy is defined as

$$S = k_B \, Ln \, \Delta\Gamma \tag{53}$$

The distribution probability function w(E) is that given by the canonical ensemble, i.e. Ln w(E) is linear with the energy. Thus one has

$$Ln \, w(E) = \alpha + \beta E \tag{54}$$

Now considering the ensemble of the subsystems, one can take the average values of different quantities. The average of Ln w(E) is

$$[Ln \, w(E)]_{av} = \alpha + \beta E_{av} = Ln \, w(E_{av}) \tag{55}$$

Since $w(E_m)\Delta\Gamma = 1$ and recalling that $E_m = E_{av}$ one gets

$$\text{Ln } \Delta\Gamma = - \text{Ln } w(E_{av}) \tag{56}$$

$$\text{Ln } \Delta\Gamma = - [\text{Ln } w(E)]_{av} \tag{57}$$

or
$$S = - k_B \sum w_n \text{ Ln } w_n = - k_B \text{ Tr}(w^o \text{ Ln } w^o) \tag{58}$$

One recovers again the preceding formula namely that the entropy is the average of the logarithm of the distribution probability.

The definition of the entropy $S = k_B \text{ Ln } \Delta\Gamma$ is equivalent to the expression (36) of Gibbs $\Phi_0 = [\text{Ln}(dV/d\mathcal{E})]_0$ with the exception of the quantity $\Delta E$ which does not appear in the Gibbs's expression. But this quantity $\Delta E$ is important since it introduces the notion of spreading of the energies.

## 7. Entropy today

The preceding sections showed us that there are several definitions of the entropy. One can put them into two classes. In the first, the entropy is a characteristic of systems with temperature i.e. with a very large number of particles. The thermodynamic version is the entropy with its differential $dS = dQ/T$. The statistical version is given by the expressions

$$S \approx - \eta_{av} = - \int P \text{ Ln } P \, dq_1.....dp_n \quad (32) \qquad S = - k_B \sum w_n \text{ Ln } w_n \quad (58)$$

The second class is related to the property of the entropy as a maximum in a closed system. The thermodynamics version states that in a process taking place adiabatically, the entropy either remains constant or increases. The statistical version is the Boltzmann formula and the expression (36) of Gibbs:

$$S \approx \text{Ln } W \quad (11) \qquad \Phi = [\text{Ln}(dV/d\mathcal{E})] \quad (36)$$

### 7.1. Entropy as a tool

Following Clausius one can see entropy as a tool: it is a function of the state of the system and in order to calculate the entropy change between two states one has to choose a reversible path. It is very useful in order to solve problems in Thermodynamics. The conjunction of the two fundamental laws of thermodynamic permits to get numerous important results that one can find in all textbooks on thermodynamics.

$$dE = dQ - PdV \quad \text{and} \quad dS = dQ/T$$

It seems that the first to follow this way was Clausius who in fact discovered the two fundamental laws.

7.2 Entropy as a potential

*An another thermodynamic potential*

The concept of thermodynamics potentials is very well known. Classically they are the Helmholtz free energy $F = U - TS$ (U is the internal energy), the Gibbs free energy $G = U - TS + PV$ (P is pressure and V the volume) and the grand potential $-PV$. When expressed as function of T, V and N (the number of particles in the system), F is a minimum for processes taking place at constant T, N. And when G is expressed as a function of T, P and N, it is a minimum for processes taking place at constant T, N. The same holds for the grand potential $\Psi$ expressed a function of T, V and the chemical potential. Now one can include the entropy in the category of the thermodynamic potentials and considers -S as a potential for processes taking place at constant energy, constant V and constant N. When one says that in process taking place in a closed system, the entropy is maximal, one can say that: in a process taking place at constant U, V and N, -S is a minimum. This gives a very simple meaning for the entropy associating it with the notion of thermodynamic potential.

To see the entropy as a potential for an isolated system permits to remove the mysterious aspect of the entropy. As in Mechanics the equilibrium is characterized by a minimum of the potential energy, in a closed system the thermodynamic equilibrium is characterized by the minimum of $-S$. As in Mechanics, the derivatives of the potential energy give the forces, the derivatives of the entropy give various thermodynamics quantities. This point of view was adopted by Callen [15] who puts the entropy on the same grounds than the Helmholtz free energy, the Gibbs free energy or the grand potential.

I shall take a very simple (even trivial) example. In spite of its simplicity, this shows how S can be seen as a potential. One considers an isolated box of volume V with an ideal gas inside and a mass hanged by a spring at the top of the box. One supposes that the heat capacity $C_G$ of the gas and that of the mass ($C_M$) are independent of the temperature. The mechanical energy of the spring is $(1/2)Kx^2$ when K is the spring constant, independent of the temperature and x is the displacement of the mass from the equilibrium position. One supposes also the internal energies to the gas and of the mass are independent of the volume: $U_G = C_G T$ and $U_M = C_M T$. The system is in equilibrium: the temperature of the gas and the mass is $T_1$ and the mass is maintained at the position $x_0$. One can liberate the mass from this position and after some time it will come to the position $x = 0$ because of the friction with the gas. The mechanical energy of the spring is transformed in heat and the temperature of the system increases.

For the sake of simplicity one can suppose that the mass will move very slowly without oscillations. One has two equations:

Conservation of the energy   $U_0 = (C_G + C_M)T + 1/2\, Kx^2$ (46)

One neglects the gravitational potential energy to simplify the problem but its inclusion gives the same conclusion.

Entropy   $S = \int (C_G + C_M)\, dT/T$ + a function of the volume + constant (58)

or   $S = (C_G + C_M) \ln T$ + constant (including the function of the volume) (59)

or   $S = (C_G + C_M) \ln[U_0 - (1/2)Kx^2]$ + constant. (60)

Now one writes that $(\partial S/\partial x) = - Kx\, (C_0 + C_M)/[U_0 - (1/2)Kx^2] = 0$. The trivial solution is $x = 0$ as expected: the mass comes to its lowest position. One can also verify that $(\partial^2 S/\partial x^2) < 0$ as necessary for a maximum.

It is interesting to compare the links relating the macroscopic quantities, entropy S, Helmholtz free energy F and grand potential $\Psi$ with the functions deduced from the microscopic properties of a system: the number of microstates W, the partition function Z and the grand partition function $Z_G$.

For a closed system one has

$$- S = - k_B \ln W \qquad (61)$$

for systems in a canonical ensemble defined by T,V and N

$$F = -k_B T \ln Z \tag{62}$$

and for systems in a grand canonical ensemble defined by T,V and the chemical potential

$$\Psi = -k_B T \ln Z_G \tag{63}$$

One sees clearly the parallelism between the three fundamental links of Statistical Mechanics

*Fluctuations*

Until now only pure equilibrium states were mentioned. However, there are fluctuations of the thermodynamic quantities around their mean or macroscopic values. An incomplete analogy can be made with mechanical equilibrium; a small perturbation will give oscillations around the equilibrium position. The theory of the thermodynamic fluctuations is exposed in several textbooks, see ref 14 and 15 and the entropy is the key for investigate fluctuations.

The starting point is the probability to "find" a certain quantity x in the interval x, x + dx when the entropy is afunction of x,S(x). This probability is proportional to $\exp[S(x)/k_B]dx$ or

$$\Pi(x)dx = C \exp[S(x)/k_B]dx \tag{64}$$

C is a normalization constant. One can also suppose that the mean value of x, <x> is subtracted from x such one can consider that the mean value of x is zero. Near the equilibrium, one can develop S(x) as

$$S(x) = S(0) + (1/2)[\partial^2 S/\partial x^2]x^2$$

or $$S(x) = S(0) - m\, x^2/2 \tag{65}$$

with $b = -[\partial^2 S/\partial x^2]$ for x = 0 and b > 0 since S(x) is maximum for x = 0.. This gives for the probability $\Pi(x)$

$$\Pi(x)dx = A \exp(-b\, x^2/2k_B)\, dx \tag{66}$$

where the constant A is given by

$$A\int_{-\infty}^{+\infty} \exp(-bx^2/2k_B)\, dx = 1 \tag{67}$$

or $$A = \sqrt{(b/k_B\, 2\pi)}$$

Finally the mean value of $x^2$ (that we call the fluctuation of x) is

$$<x^2> = \sqrt{(b/k_B\, 2\pi)} \int_{-\infty}^{+\infty} x^2 \exp(-bx^2/2k_B)\, dx \qquad (68)$$

or
$$<x^2> = (k_B/b) \qquad (69)$$

We shall calculate the fluctuations of the position of the mass in the preceding example of a spring in contact with a gas i.e. we shall determine the fluctuation of x, $<x^2>$. From (65) one has

$$b = -[\partial^2 S/\partial S^2]_{x=0} = (C_0 + C_M)\, K/U_0 \qquad (70)$$

from which one deduces

$$<x^2> = k_B\, U_0/[(C_0 + C_M)\, K] \qquad (71)$$

At the final equilibrium, $U_0 = (C_0 + C_M)\, T_f$ ($T_f$ is the final temperature), one can write $<x^2>$ as

$$<x^2> = k_B\, T_f/K \qquad (72)$$

which is a standard result of the thermodynamic fluctuations theory.

*Entropy out of equilibrieum*

Landau and Lifshitz [14] propose a definition of the entropy for states out of equilibrium. Suppose that the system with relaxation time $\Delta t$ is not in an equilibrium state. One can divide (a thought experiment) the system into numerous small (but macroscopic) subsystems with relaxation time much smaller than $\Delta t$. It is possible to define for each these subsystem its entropy and the total entropy is the sum of all these partial entropies.

Furthermore there is a long tradition among those who use the microcanonical ensemble to calculate the entropy via the formula $S = k_B\, \text{Ln}\, W$. The quantity W can be well defined even for states out of equilibrium and the final stage is to find the conditions for W to be a maximum. It is the standard method exposed in several textbooks to obtain the Maxwell-Boltzmann, the Fermi-Dirac and the Bose-Einstein distributions.

7.3. Entropy and irreversibility.

About irreversibility, the work of Planck [18] is well known. For him, the notion of process, in particular irreversible process, is essential for the definition of the entropy:

*If we regard the second law from the mathematical point of view, the distinction between the final and initial states of a process can consist only in an inequality. This means that a certain quantity, which depends on the momentary state of the system, possesses in the final state a greater or smaller value, according to the definition of the sign of that quantity, then in the initial state.*

Furthermore irreversibility [16] is not a simple concept and needs special attention and clarification. We take a very simple example of the free expansion of gas in a close vessel. In the final state the entropy is larger than in the initial state. It is often considered that this is the signature of an irreversible process. But one can see the things in another manner. The entropy is a state function and becomes larger because the parameters of the final state are such that the entropy increases. The fact it is possible to pass also from the initial state to the final state in a reversible way could be an indication that the increase of the entropy is not necessarily related to an irreversible process. This is why Planck proposes his own definition of irreversible process. A process is irreversible if, in order to come back to the initial state of the system, it is necessary to induce a change in the state of environment of the system. In the case of the free expansion, in order to come back to the initial state of the gas, first the volume of the gas must be decreased doing work from the environment and then the temperature of the gas must be brought to its initial value by thermal contact with the a reservoir. Finally the environment did work on the gas and exchanged heat with the gas. In other words, in the complete process of free expansion of a gas and its return to the initial state, the environment did not return to its initial state.

Introducing the entropy through irreversible processes is a very standard method to define the entropy. However this method presents some difficulties. First, because in the development of Thermodynamics, one uses derivatives and integrals and this is possible only if all the states of a system are equilibrium states using the concept of reversible quasi-static processes. In this part of Thermodynamics called "Thermostatics" (in contrast to Thermodynamics of irreversible process) one never considers irreversible processes. There is some paradox to introduce entropy using the concept of irreversibility to study phenomena in which irreversible processes never appear. Another paradox is as follows: in order to calculate the entropy which increases in an irreversible process, one has to find a reversible path between the initial and the final states. Reversible processes are necessary to characterize irreversible processes!

# 8. Conclusion

In this trip towards entropy, I tried to find aspects of the entropy which are relatively easy to understand: entropy as a tool, entropy as a potential, entropy as indication of distribution and fluctuations, in the microcanonical ensemble through the Boltzmann formula and in the canonical ensemble through the Gibbs formula, entropy as the gate to the Quantum Mechanics.

The property of the entropy to be a maximum in some circumstances has always excited the imagination.. Already from the beginning, there is the famous sentence of Clausius, (the last line of his text):

*The entropy of the universe tends to a maximum*

I wonder about the impact of this notion of a maximum. What is this particular quantity which can only increase? On other hand, it appears so "natural" to associate an equilibrium state with a minimum of some quantity like the potential energy that associating it with a maximum seems a paradox. May be a part of the mystery comes from this association of equilibrium and maximum of entropy.

The intriguing aspect of entropy is, above all, due to the richness of the concept. In this trip toward the past, it was not possible to discuss all the facets of the concept nor the different methods to expose the Second Law. There are numerous works on these other aspects and I refer the reader to ref.16. Most recent developments (Tsallis) are also absent since my intention was to investigate the emergence of the concept.

## Appendix A

<u>Derivation of the expression</u>  $d\mathcal{E}_{av} = -\Theta \, d\eta_{av} - \Sigma \, (A_i)_{av} \, da_i$

One begins with (29)

$$\exp(-\Psi/\Theta) = \int \exp(-\mathcal{E}/\Theta) \, dq_1\ldots dq_n dp_1\ldots dp_n \qquad (29)$$

and derives the both sides relatively to $\Theta$

$$\exp(-\Psi/\Theta)(-d\Psi/\Theta + \Psi \, d\Theta/\Theta^2) = (d\Theta/\Theta^2)\int \mathcal{E} \exp(-\mathcal{E}/\Theta) \, dq_1\ldots dq_n dp_1\ldots dp_n$$

$$-1/\Theta \, \Sigma \, da_i \int (d\mathcal{E}/da_i) \exp(-\mathcal{E}/\Theta) \, dq_1\ldots dq_n dp_1\ldots dp_n \qquad (1A)$$

Multiplying both sides by $\Theta \exp(-\Psi/\Theta)$ and rearranging the terms gives

$$-d\Psi = \Psi \, d\Theta/\Theta + d\Theta/\Theta \int \mathcal{E} \exp[(\Psi - \mathcal{E})/\Theta] \, dq_1\ldots dq_n dp_1\ldots dp_n$$

$$-\Sigma \, da_i \int (d\mathcal{E}/da_i) \exp[(\Psi - \mathcal{E})/\Theta] \, dq_1\ldots dq_n dp_1\ldots dp_n \qquad (2A)$$

Taking into account that $d\mathcal{E}/da_i = A_i$ and that averages are given by (28), one gets

$$d\Psi = [(\Psi - \mathcal{E}_{av})/\Theta] \, d\Theta - \Sigma \, (A_i)_{av} \, da_i \qquad (3A)$$

But $[(\Psi - \mathcal{E}_{av})/\Theta] = \eta_{av}$ and $d\Psi = d\mathcal{E}_{av} + \eta_{av} \, d\Theta + \Theta \, d\eta_{av}$. Putting in (3A) gives the desired expression

$$d\mathcal{E}_{av} = -\Theta \, d\eta_{av} - \Sigma \, (A_i)_{av} \, da_i$$

## Appendix B

<u>Calculation of $d\Phi/d\mathcal{E}$</u>

First we calculate the average value of $d\Phi/d\mathcal{E}$. Since $\Phi = \text{Ln}(dV/d\underline{\mathcal{E}})$, one has

$$d\Phi/d\varepsilon = (d^2V/d\varepsilon^2)/(dV/d\varepsilon) \tag{B1}$$

and taking into account (34)

$$(d\Phi/d\varepsilon)_{av} = \int (d\Phi/d\varepsilon)\,\exp[(\Psi - \varepsilon)/\Theta]\,(dV/d\varepsilon)d\varepsilon \tag{B2}$$

Inserting (B1) in (B2) gives

$$(d\Phi/d\varepsilon)_{av} = \int \exp[(\Psi - \varepsilon)/\Theta]\,(d^2V/d\varepsilon^2)d\varepsilon \tag{B3}$$

An integration by parts gives

$$(d\Phi/d\varepsilon)_{av} = G(\text{smallest }\varepsilon) - G(\varepsilon = \infty) + (1/\Theta)\int \exp[(\Psi - \varepsilon)/\Theta]\,(dV/d\varepsilon)d\varepsilon \tag{B4}$$

with $G(\varepsilon) = (dV/d\varepsilon)\exp[(\Psi - \varepsilon)/\Theta]$. Gibbs shows that $(dV/d\varepsilon)$ goes to zero for the smallest value of the energy and clearly G goes to zero for the energy going to infinity.. By definition of a probability, the integral of the right side is equal to 1 (see (34)).

It results that $(d\Phi/d\varepsilon)_{av} = 1/\Theta$.

Now we calculate the value of $d\Phi/d\varepsilon$ for the most probable energy. Consider the quantity

$$Y = N\exp[(\Psi - \varepsilon)/\Theta]\,(dV/d\varepsilon) = N\exp[\Phi + (\Psi - \varepsilon)/\Theta] \tag{B5}$$

The last equality comes from the fact that $\exp\Phi = dV/d\varepsilon$, by definition of $\Phi$. N is the number of system is the ensemble. As mentioned above, the function Y goes to zero for the smallest value of the energy and for the energy going to infinity. Consequently the function has a maximum. Y gives the number of systems in the ensemble having energy between $\varepsilon$ and $\varepsilon + d\varepsilon$ and the maximum of Y gives the most probable energy. The maximum is given by $dY/d\varepsilon = 0$.

$$dY/d\varepsilon = N(-1/\Theta + d\Phi/d\varepsilon)\exp[\Phi + (\Psi - \varepsilon)/\Theta] = 0 \tag{B6}$$

$dY/d\varepsilon = 0$ if $d\Phi/d\varepsilon = 1/\Theta$. In other words, $d\Phi/d\varepsilon = 1/\Theta$ for the most probable value of the energy.

## Appendix C

Ln V as the entropy

We calculate the average of $d\mathcal{E}/d(\ln V)$ to show that Ln V is also a candidate for the entropy.

$$[d\mathcal{E}/d(\ln V)]_{av} = \int d\mathcal{E}/d(\ln V)\ \exp[(\Psi - \mathcal{E})/\Theta]\ (dV/d\mathcal{E})d\mathcal{E} \tag{C1}$$

Since $d(\text{Ln } V) = (dV/d\mathcal{E})(d\mathcal{E}/V)$, (C1) becomes

$$[d\mathcal{E}/d(\ln V)]_{av} = \int \exp[(\Psi - \mathcal{E})/\Theta]\ V\ d\mathcal{E} \tag{C2}$$

An integration by parts gives

$$d\mathcal{E}/d(\ln V)]_{av} = F(\text{smallest } \mathcal{E}) - T(\mathcal{E} = \infty) + \Theta \int \exp[(\Psi - \mathcal{E})/\Theta]\ (dV/d\mathcal{E})d\mathcal{E} \tag{C3}$$

with $F(\mathcal{E}) = -V\Theta \exp[(\Psi - \mathcal{E})/\Theta]$. The function $F(\mathcal{E})$ is null for the smallest of the energy since as $V = 0$ and it is also null for the energy going to infinity. Consequently

$$d\mathcal{E}/d(\ln V)]_{av} = \Theta \tag{C4}$$

since $\int \exp[(\Psi - \mathcal{E})/\Theta]\ (dV/d\mathcal{E})d\mathcal{E} = 1$ (see (34))

Furthermore, it not very difficult to see that the average of $\exp(-\Phi)V$ is given by

$\int \exp[(\Psi - \mathcal{E})/\Theta]\ V\ d\mathcal{E}$. And we showed above that this integral is equal to $\Theta$.